# Modeling the Slow Arrhenius Process (SAP) in Polymers

Valeriy V. Ginzburg,[1,*] Oleg V. Gendelman,[2] Simone Napolitano,[3] Riccardo Casalini,[4] and Alessio Zaccone[5]

[1]Department of Chemical Engineering and Materials Science, Michigan State University, East Lansing, Michigan, USA 48824

[2]Faculty of Mechanical Engineering, Technion, Haifa 3200003, Israel

[3]Laboratory of Polymer and Soft Matter Dynamics, Experimental Soft Matter and Thermal Physics (EST), Universite libre de Bruxelles (ULB), Brussels 1050, Belgium

[4]Chemistry Division, Naval Research Laboratory, 4555 Overlook Avenue SW, Washington, D.C., USA 20375

[5]University of Milan, Department of Physics, via Celoria 16, 20133 Milano, Italy

*Corresponding author: ginzbur7@msu.edu

**Abstract**

Amorphous glass-forming polymers exhibit multiple relaxation processes, including the structural α-relaxation associated with the glass transition and faster secondary relaxations that typically follow Arrhenius behavior. Recently, a distinct slow Arrhenius process (SAP) has been observed at frequencies well below the α-process. Although Arrhenian in its temperature dependence, the SAP involves much longer relaxation times and its microscopic origin remains unclear. Here, we extend the two-state, two-timescale (TS2) theory to describe both the α-relaxation and the SAP within a unified framework. We propose that the SAP represents the high-temperature limit of an α-like process in a coarse-grained fluid of dynamically correlated clusters. With renormalized interaction energies and coordination parameters, the same model quantitatively reproduces both α and SAP data across multiple polymers without additional adjustable parameters and explains the observed Meyer–Neldel compensation behavior. The theory further predicts that the SAP should deviate from Arrhenius behavior at sufficiently low temperatures, transitioning to Vogel–Fulcher–Tammann–Hesse-like dynamics, thereby offering a physically transparent interpretation of cluster-scale relaxation in glass-forming polymers.

## 1. Introduction

The recent experimental discovery of the slow Arrhenius process (SAP) resulted in a significant change in our understanding of the near-glassy dynamics of amorphous polymers.[1–6] The SAP, first observed in the broadband dielectric spectroscopy (BDS)[7,8] experiments at frequencies significantly below those of the well-known $\alpha$, $\beta$, and other transitions, has now also been reported by a range of complementary experimental techniques[5,9] and simulations.[10] As a function of temperature, this process exhibits Arrhenius behavior, and the activation energy and the prefactor are well-correlated via the "entropy-enthalpy compensation" or "Meyer-Neldel rules"[3,5]. The SAP is thought to play an important part in determining the dynamics of various processes, from polymer adsorption[5] to crystal growth in small organic molecules[11] and lipid exchange in membranes.[6]

Theoretical understanding of the SAP is still being developed. White, Napolitano, and Lipson (WNL)[4] formulated the so-called Collective Small Displacement (CSD) model. Within their framework, SAP corresponds to the reshaping of large clusters of molecular segments, unlike the $\alpha$-transition, which is associated with the cage-breaking by individual atoms or repeat units. Lipson and co-workers[4,6] demonstrated that their model predicts the Arrhenius dependence of relaxation time on temperature and that the activation energy is correlated to the van-der-Waals-type energy parameter in their "locally correlated lattice" (LCL) equation of state.[12–14] While the thermodynamic equation (LCL) is equally applicable to the $\alpha$ (glass transition) and the slow Arrhenius processes, the dynamics of the former are described by a different model, called Cooperative Free Volume (CFV).[15–19] Within CFV, the relaxation time dependence on temperature is strongly non-Arrhenius, consistent with experimental data and the well-known ideas of Adam and Gibbs,[20] Vogel, Fulcher, Tammann and Hesse,[21–23] Doolittle,[24] KSZ,[47] and others.

The SAP activation energy is close to the high-temperature viscosity thermal barrier. For many glass-forming materials, the β process is Arrhenius and corresponds to the Johari-Goldstein (JG)[25,26] secondary relaxation. Furthermore, in many of these materials, the JG β-process merges with the α–process[27] at some finite temperature $T_{\alpha\beta}$ (also called the Arrhenius temperature or $T_A$, see below). The combined relaxation process at temperatures $T > T_{\alpha\beta}$ is sometimes labeled the αβ-process, is Arrhenius, and is related to the shear viscosity of the material in the liquid state (e.g., polymer melts). In Arrhenius coordinates, the SAP line is parallel to the high-temperature αβ-line.[5] Thus, in principle, a comprehensive theory of SAP could be derived from the description of those two processes.

Recently, we have formulated the so-called "two-state, two-timescale" (TS2) theory describing the α- and β-processes in conventional glass-formers (network, molecular, and polymeric).[28–34] To better relate the model to experimental dilatometry data, the TS2 model was combined with the Sanchez-Lacombe (SL)[35–38] equation of state. Interestingly, although SL-TS2 model requires five parameters to describe the relaxation time dependence on the temperature (at a constant pressure), we found that only two of those quantities are material-dependent (the rest are universal).[34] In other words, with just two parameters (e.g., $T_g$ and fragility), we can recover the whole equilibrium relaxation time vs. temperature curve for the α- and β-processes. We now ask – can this approach be leveraged to include SAP?

Our starting point is the assumption of White, Napolitano, and Lipson[4] that SAP is associated with the dynamics of large clusters, rather than individual particles. We assume that the clusters can be considered as "coarse-grained particles" (CGP), on the relevant time-scale window. The new "coarse-grained" material (CGM) would be still a conventional glass-former, but with the new (rescaled) "general TS2" parameters. The SAP in that case would correspond to the $\alpha\beta$-process for the CGM, and the glass-liquid thermodynamic transition temperature for the CGM would be slightly lower than for the material itself. Below, we will test this assumption against experimental data for multiple polymers. We will also see how the enthalpy-entropy compensation rules emerge within this new framework. Finally, we will discuss remaining questions and potential applications of the new theory.

Importantly, the clusters considered here are dynamically emergent entities defined by the separation of timescales between intra-domain and inter-domain relaxation. Within the window $\tau_\beta \ll t \ll \tau_{SAP}$, these clusters behave as effectively rigid units, allowing the SAP to be interpreted as the structural relaxation of a renormalized coarse-grained fluid.

## 2. The Model

### 2.1. The General TS2 Modeling and Parameterization of the $\alpha$-Transition

We start by briefly summarizing the "two-state, two-(time) scale" (TS2) model. Within TS2, it is assumed that in amorphous materials, there are "domains" in which all the atomic motions are strongly correlated. Those domains can be thought of, e.g., as an atom (or repeat unit) and two or three of its coordination shells. (In the past, we borrowed the term "cooperatively rearranging region" (CRR)[20] to describe those domains; however, our domains are probably best described as the "high-temperature limit CRR"). The correlation length, $\xi$, which is the cubic root of the CRR volume, depends weakly on the temperature due to the formation of large clusters of glassy or immobile domains).[39] Each domain can be either in the "Liquid" (1) or the "Solid" (2) state.

Within TS2, we concentrate on the $\alpha$- and $\beta$-transitions. The "fast" Johari-Goldstein (JG)[25,26] $\beta$-relaxation time is given by the Arrhenius function,

$$\tau_\beta[T] = \tau_{el} \exp\left[\frac{E_1}{k_B T}\right] \tag{1}$$

The "slow" $\alpha$-relaxation time is a super-Arrhenius function, increasing sharply as the temperature approaches the glass transition temperature, $T_g$, from above. This behavior is phenomenologically described by many different expressions (Vogel,[21] Fulcher,[22] Tammann and Hesse,[23] Williams, Landel, and Ferry,[40] Avramov and Milchev,[41] Mauro and co-workers,[42] KSZ,[43] and others). Here, we use the expression proposed by Ginzburg in his TS2 model,[28]

$$\tau_\alpha[T] = \tau_{el} \exp\left[\frac{E_1}{k_B T} + \frac{E_2 - E_1}{k_B T}\psi\right] \tag{2}$$

In eqs 1-2, $E_1$ and $E_2$ are the activation energies in the liquid and solid states, respectively; $\tau_{el}$ is the "elementary" (high-temperature-limit) material time, and $\psi$ is the solid fraction (i.e., the fraction of domains in the solid state at any given time), described as,

$$\psi = \frac{1}{1 + \exp\left[\Delta S\left(1 - \frac{T_X}{T}\right)\right]} \tag{3}$$

Here, $T_x$ is the thermodynamic glass-liquid transition temperature (not to be confused with the dynamic glass transition temperature, $T_g$), and $\Delta S$ is the non-dimensional entropy difference between the two states. Equation 3 emerges naturally from the standard Boltzmann framework applied to a two-state system. The TS2 model thus has five parameters. Recently, we have shown[34] that out of these five parameters, three are universal (when properly converted to a non-dimensional form), and only two ($T_x$ and $\tau_{el}$) are material specific. (Note that the current analysis applies to an isobaric experiment where pressure is kept constant ("atmospheric"); performing an isobaric experiment at a different pressure would mean new values of $T_x$ and $\tau_{el}$).

The general TS2 equations can be re-written in the following form,

$$\tau_\beta[\overline{T}] = \tau_{el} \exp\left[\frac{\overline{E}_1}{\overline{T}}\right] \tag{4a}$$

$$\tau_\alpha[\overline{T}] = \tau_{el} \exp\left[\frac{\overline{E}_1}{\overline{T}} + \frac{\overline{E}_2 - \overline{E}_1}{\overline{T}}\psi[\overline{T}]\right] \tag{4b}$$

$$\psi[\overline{T}] = \frac{1}{1 + \exp\left[\Delta S\left(1 - \overline{T}^{-1}\right)\right]} \tag{4c}$$

$$\overline{T} = \frac{T}{T_X} \tag{4d}$$

Here, the three universal parameters are: $\overline{E}_1 = 50$, $\overline{E}_2 = 150$, and $\Delta S = 16$. As discussed in our recent paper,[34] equations 4a—4c can describe a broad family of curves for various polymeric, molecular, and network glass-formers.

It is convenient to rewrite eq 4a—4d in terms of the so-called Arrhenius temperature, $T_A$, and Arrhenius time, $\tau_A$. Following Hung et al.,[44] we define $T_A$ as the temperature where the deviation of the α-relaxation time from the high-temperature Arrhenius asymptotic equals 5%, and $\tau_A$ as the α-relaxation time at T = $T_A$. The temperature dependence of the α-relaxation time can then be written as,

$$\log\left(\frac{\tau_\alpha[\overline{T}]}{\tau_A}\right) = \left(\frac{1}{\ln(10)}\right)\left[\left(\frac{\overline{E_1}}{\overline{T}} - \frac{\overline{E_1}}{\overline{T_A}}\right) + \frac{\overline{E_2} - \overline{E_1}}{\overline{T}}\psi[\overline{T}]\right]$$

(5a)

$$\overline{T_A} = \frac{T_A}{T_X}$$
(5b)

It turns out that the value of $\overline{T_A}$ is material-independent, $\overline{T_A} \approx 1.23$,[34] consistent with empirical observations.[45] Thus, we will subsequently use $T_A$, rather than $T_X$, to represent the temperature scale of a material. Likewise, the relationship between $\tau_A$ and $\tau_{el}$ is material-independent, $\log(\tau_A) = \log(\tau_{el}) + 17.7$.

The above analysis has been purely phenomenological. A more physical interpretation of TS2 has been proposed by Ginzburg et al.[29–32] in several subsequent papers. The new framework combines the TS2 dynamics with a thermodynamic equation of state (specifically, the Sanchez-Lacombe[35–37] lattice model). In this model, the liquid state is approximated as a flexible chain with $r_L$ "monomers", and the solid state is approximated as a chain with $r_S$ ($r_S < r_L$) "monomers". In addition, some lattice sites are taken by "voids" or "free volume elements" – the void volume fraction or the "free volume" is labeled (1-$\nu$), where $\nu$ is the "occupancy". The nearest neighbors interact via local "physical bonding" (non-covalent) interactions characterized by interaction energies $\varepsilon_{ij}$, $i, j = L, S$. This analysis helps us describe the dielectric relaxation and the dilatometry experiments within a single model. It can also explain the approximate universality of the ratio $\alpha_L/\alpha_G \approx 3$[46–48] (where $\alpha_L$ and $\alpha_G$ are the volumetric coefficients of thermal expansion) and the "Boyer rules" for polymers.[49–51] The transition between the solid and liquid states happens on the shorter timescale ($\tau_\beta$), while the creation and destruction of the voids happens on the longer timescale ($\tau_\alpha$). For more details, see Ginzburg et al.[29,31]

### 2.2. Application of the TS2 Model to Slow Arrhenius Process

Our starting assumption is that SAP is not an exotic phenomenon, but a ubiquitous process inherently present in most if not all glass-forming materials. In that case, its theoretical description should also rely on some simple and universal mechanisms. As discussed above, White, Napolitano, and Lipson (WNL)[4] recently formulated the "collective small displacements" model that captured many features of the SAP, including the Meyer-Neldel[52] compensation rules (the relationship between the slopes and intercepts of the SAP curves, in Arrhenius coordinates, for different materials). The model discussed here is based on a different physical mechanism, although the resulting phenomenology is in many ways similar to that of WNL.

The main idea of our new model is visualized in Figure 1. In our formalism, the $\alpha$– (and $\beta$–) processes are related to the motions on the molecular level ("within one domain"). We consider monomers with binding energy $\varepsilon_\alpha$ (for simplicity, we disregard the differences between $\varepsilon_{\alpha,SS}$, $\varepsilon_{\alpha,LS}$, and $\varepsilon_{\alpha,LL}$, which are on the order of several percent) and the coordination number $Z_\alpha$. We also hypothesize that the neighboring particles can interact via "strong" (labeled as orange) or "weak" (labeled as grey) interactions. The coordination number counts only the strong bonds – thus, e.g., in Figure 1a, $Z_\alpha$ = 3.7.

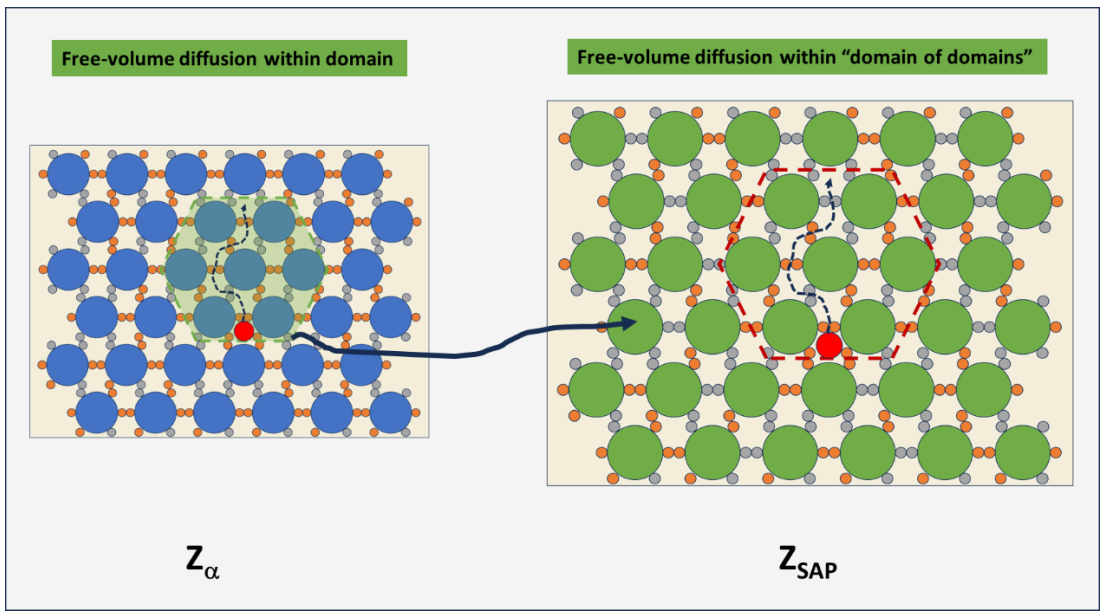

*Figure 1. Left panel – small scale ($\alpha$-process). The blue circles are monomers or molecules. The orange circles correspond to "active" connections, and the grey circles correspond to the "passive" ones. The green dashed line and the green fill highlight the boundaries of a single domain. The red circle is a void diffusing through the domain, and the black line shows its trajectory. Right panel – large scale (SAP). The green circles are the domains. The meaning of the orange, grey, and red circles is similar to that in the left panel.*

Now, the cooperativity within the domains implies that one can coarse grain to consider one domain to be a "coarse-grained particle" (CGP). This is visualized on the right panel in Figure 1. After such a transformation, the new fluid of CGPs must obey the same thermodynamics and dynamics as the original fluid of monomers, albeit with different parameters (energy scale, coordination number, and elementary time). We expect that in general, the coordination between CGPs is smaller than between molecular fragments, thus, e.g., in Figure 1b, $Z_{SAP}$ = 3.0.

Our crucial assumption is that, in the CGP fluid, the SAP effectively replaces the α-relaxation. Furthermore, given that experiments and simulations shows that the SAP is an ***Arrhenius*** process, we expect that the thermal barrier of the SAP should be temperature invariant in the region corresponding to T > $T_{A,SAP}$, where $T_{A,SAP}$ is the "Arrhenius temperature" of the CGP fluid (see the definition in the previous section and also Hung et al.)[44] Above $T_{A,SAP}$, there are almost no solid domains ($\psi \approx 0$) and the α- and β-transitions coincide. One expects that – if our assumptions are correct – the SAP relaxation time could eventually become non-Arrhenian as a function of temperature if one continues measurements at low temperatures.

Thus, we will apply the general TS2 model (eqs 4a – 4d) to describe the SAP relaxation time without any modifications. The SAP activation energy, $E_{SAP}$, is consequently $E_{1,SAP}$, and the transition temperature $T_{X,SAP} = E_{SAP} / \left( R \overline{E}_1 \right)$. Likewise, the Arrhenius intercept in the SAP fitting should correspond to $\log(\tau_{el,SAP})$. To test such ideas, in the next section, we will see how the SAP and α-relaxation parameters compare for each individual polymer and across different polymers.

### 2.3. The Meyer-Neldel Compensation Law for the Slow Arrhenius Process

Napolitano and co-workers[3,4] have shown that for most SAP data, the Meyer-Neldel compensation law (CL)[52,53] holds, i.e., given the Arrhenius temperature dependence of the SAP relaxation time,

$$\ln(\tau_{SAP}[T]) = \ln(\tau_{A,SAP}) + \left[\frac{E_{a,SAP}}{k_B T} - \frac{E_{a,SAP}}{k_B T_{A,SAP}}\right] \tag{6a}$$

there is a linear relationship between the slope and the intercept,

$$\ln(\tau_{A,SAP}) = A - k E_{a,SAP} \tag{6b}$$

In other words, the materials that exhibit SAP differ only by their activation energy, $E_{a,SAP}$. This is surprising in a sense that in general, an arbitrary set of Arrhenius materials should be a scatter of points in the ($\tau_{A, SAP}$, $E_a$) plane; in reality, as observed for other physical phenomena[54–56] they all sit on a single line, as manifested by eq 6b.[3–5] What is the cause of this unexpected behavior?

The White-Napolitano-Lipson (WNL) framework suggested that the origin of the SAP data collapse onto a single line is in the existence of the "material-independent baseline energetic contribution".[4] This same trend can be derived within a different physical picture in the context of TS2 approach. Within this formalism,[29,31] one can write,

$$T_X = a\frac{\varepsilon^*}{k_B} = a\frac{Z\varepsilon_{SS}}{2k_B} \tag{7}$$

Here, $a \approx 0.4$ is a universal parameter within SL-TS2, $\varepsilon_{ss}$ is the interaction energy between the nearest neighbors in the solid/glassy state, and $Z$ is the coordination number (see below for more detailed discussion). For the SAP activation energy, we thus have,

$$E_a = \overline{E}_1 T_{X,SAP} = a \frac{Z_{SAP} \varepsilon_{SAP}}{2k_B} \overline{E}_1 \qquad (8)$$

Here, $Z_{SAP}$ is the coordination number for the "coarse-grained fluid" (CGF) and $\varepsilon_{SAP}$ is the interaction energy of the solid/glassy state of CGF.

Based on eq 8, we can re-write eq 6b as,

$$\ln(\tau_{A,SAP}) = A - \overline{E}_1 k a \frac{\varepsilon_{SAP}}{2k_B} Z_{SAP} \qquad (9)$$

Now, let us consider this carefully. First, let us note that $\overline{E}_1$ is the ratio of the activation energy to the enthalpy (per particle) of the phase transition. Thus, we can re-interpret it as the number of particles (monomer units for the alpha transition; domains for the SAP) that a void need to "pass" en route to move through a domain (for the $\alpha$-transition) or a "domain of domains" (for the SAP). Thus, let us re-label $\overline{E}_1$ as $N_0$ to re-emphasize this point. Next, let us note that the parameter $k$ must have the unit of inverse energy. Thus, we can stipulate that $k \propto (\varepsilon_{SAP})^{-1}$, and, similarly to CSD, that $\varepsilon_{SAP}$ is material independent. In that case, we can write,

$$\ln(\tau_{A,SAP}) = A - bN_{SAP} \qquad (10)$$

Here, $b$ is a new universal dimensionless coefficient, and $N_{SAP} = N_0 \dfrac{Z_{SAP}}{2}$ is the number of bonds that present themselves as obstacles to a CGF void as it diffuses through the material. The values of $A$ and $b$ will be determined in the next section.

Equation 10 seems counterintuitive: increasing the number of bonded neighbors, Z, leads to a shorter relaxation time. In other words, a more rigid material appears to relax faster at the elementary level. A possible explanation is sketched in Figure 2.

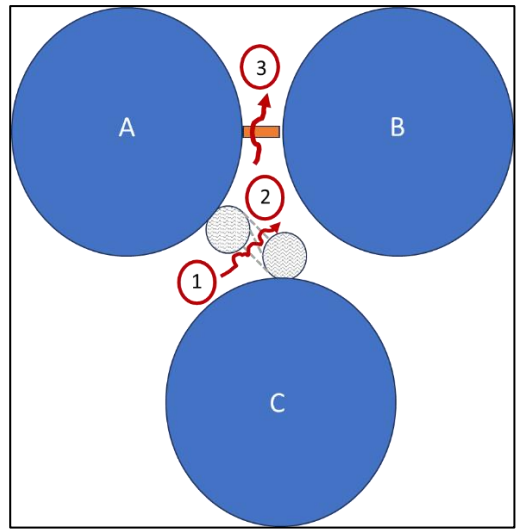

Figure 2. Sketch describing the relaxation time estimates for a void moving between two bonded or two non-bonded particles.

Consider a scenario where a "void" (or a small particle) attempts to diffuse through a narrow space between two large particles. If the two particles are bonded (like particles A and B, with the orange line representing a bond), everything is fixed, and the void navigates its way around the obstacle (red line from position 2 to position 3). Yet if the two particles are not bonded (like particles A and C), the moieties that could have bonded (two grey circles) are not completely fixed and move around in space. As a result, the trajectory of the void becomes more chaotic (red line from position 1 to position 2), and the passage time increases correspondingly. Thus, we can say that when two particles are not bonded, a void has to wait longer to diffuse past them as compared to when they are bonded. If the ratio of these two characteristic times is labeled $\lambda$, then eq 12 is equivalent to,

$$\tau_{A,SAP} = \tau^* \exp\left[b\left(N_{max} - N_{SAP}\right)\right] = \tau^* \left[\exp(b)\right]^{N_{max} - N_{SAP}} = \tau^* \lambda^{N_{max} - N_{SAP}} \quad (11)$$

In other words, the "unrealized" potential bonds, $N_{max} - N_{SAP}$, force the diffusing void to spend significant extra time navigating them. The lower-bound estimate for the Arrhenius time, $\tau^*$, was found to be on the order of $10^{-11}$ s.[34] We will apply this analysis to the data in the next section.

## 3. Results and Discussion

### *3.1. Results*

In Table 1, we compile the regressed TS2 model parameters for the $\alpha$- and SAP relaxation times, based on the data from Thoms and Napolitano.[3] As discussed above, TS2 parameterization has two free parameters for each material, thus the Arrhenius parameters can be uniquely transformed into TS2 parameters. Specifically, if the SAP is described by the Arrhenius relationship between the peak frequency and the absolute temperature,

$$\ln(f) = \ln(f_0) - \frac{E_{SAP}}{RT} \quad (12)$$

then the TS2 parameters of the SAP are given by,

$$\log(\tau_{el,SAP}) = -\frac{1}{\ln(10)} \ln(2\pi f_0) \qquad (13a)$$

$$T_{X,SAP} = \frac{E_{SAP}}{50R} \qquad (13b)$$

For the α-process, the experimental data were digitized and fit to the TS2 model using Excel Generalized Reduced Gradient (GRG) approach. For convenience, we show $\tau_A$ and $T_A$ instead of $\tau_{el}$ and $T_x$ – the relationship between those parameters is given by, $T_A = 1.23 T_X$ and $\log(\tau_{el}) = \log(\tau_A) - 17.7$.[34]

*Table 1. TS2 model parameters for α-process and SAP.*

| Material | α-process | | | SAP | | |
|---|---|---|---|---|---|---|
| | $T_A$, K | $\log(\tau_A, s)$ | $T_g$, K | $T_A$, K | $\log(\tau_A, s)$ | $T_g$, K |
| P4ClS | 437 | -5.9 | 396 | 358 | 2.1 | 376 |
| P4MS | 406 | -5.6 | 370 | 272 | 5.4 | 337 |
| PBzMA | 358 | -4.4 | 330 | 258 | 5.6 | 330 |
| PC | 465 | -8.3 | 412 | 272 | 5.5 | 340 |
| PEMA | 367 | -5.2 | 335 | 271 | 5.5 | 339 |
| PMMA | 412 | -4.8 | 378 | 355 | 2.3 | 375 |
| PNPMA | 326 | -2.7 | 307 | 337 | 2.6 | 363 |
| PS | 405 | -5.9 | 367 | 268 | 5.6 | 337 |
| PTBMA | 424 | -5.2 | 387 | 271 | 5.2 | 331 |
| PTBS | 427 | -2.5 | 404 | 279 | 5.3 | 345 |
| PTBuA | 339 | -5.3 | 310 | 258 | 6.1 | 335 |
| PVAc | 326 | -3.9 | 303 | 187 | 9.0 | 310 |
| PIB | 195 | 0.6 | 198 | 243 | 6.4 | 323 |

The thirteen polymers studied here have a broad range of glass transition temperatures and fragilities. We see that for each polymer (except PIB and PNPMA), $T_{A,SAP} < T_{A,\alpha}$, and $\log(\tau_{A,SAP}) > \log(\tau_{A,\alpha})$. We also estimated the "SAP glass transition temperature", $T_{g,SAP}$, defined as the point where $\log(\tau_{SAP}) = 2$.

Figure 3 depicts the experimental data and TS2 model fits for polystyrene (PS). The blue circles represent the experimental data for the α-relaxation, and the orange circles represent the data for the SAP. The blue and orange lines correspond to the TS2 fits for α-process and SAP. The green line is the TS2 estimate for the Johari-Goldstein β-process (the α- and β-processes merge at high temperatures).

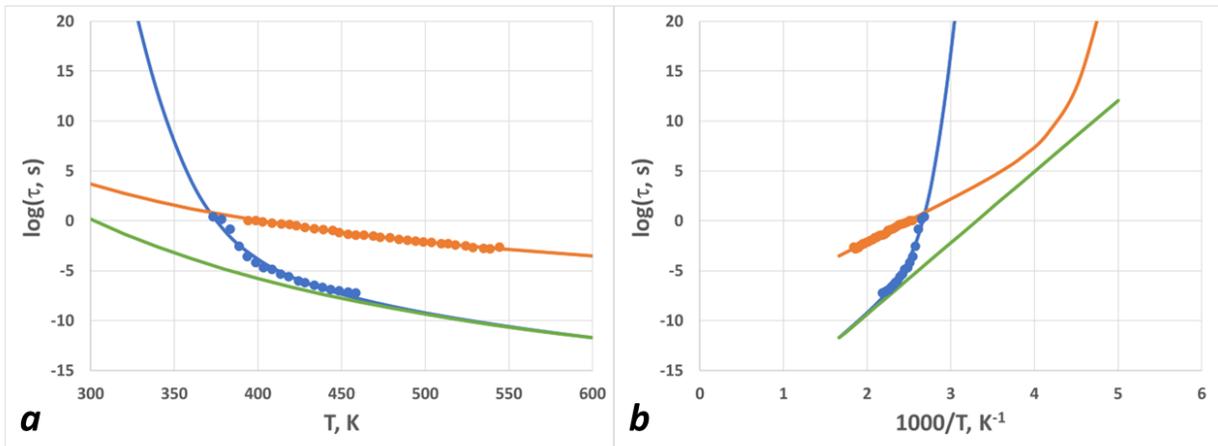

Figure 3. (a) Experimental data (circles) and TS2 model fits for PS. (b) Same as (a), but in Arrhenius coordinates. Orange circles are the data for SAP relaxation times, and blue circles are the data for the α-relaxation times. Orange line is the TS2 model for SAP, blue line is the TS2 model for the α-process, and green line is the SAP model for the JG β-process.

The glass transition of PS is approximately 367 ± 5 K, so if we consider the behavior of the material near and above $T_g$, the SAP curve is Arrhenius and nearly parallel to the TS2 JG line. The JG line continues to high temperatures as the αβ-line, and its activation energy is comparable to the flow activation energy in the melt state. All this is consistent with the earlier analysis of Napolitano et al.[3,4] We need to extrapolate to significantly lower temperatures or higher inverse temperatures (Figure 3b) to see the new features. The model predicts that the SAP line will exhibit a transition from Arrhenius to VFTH behavior at temperatures near 250 K (1000/T = 4.0), or about 100 K below the glass transition. The "SAP glass transition temperature" is estimated to be 337 ± 5 K. Although the experimental signature of this transition is still unknown, it would be interesting to see whether this transition can be picked up in experiments such as Differential Scanning Calorimetry (DSC), dilatometry, or any other. The effect, however, might be difficult to observe if the deviations from the Arrhenius behavior occur at very long times/small frequencies.

The above analysis was for the relatively "fragile" polystyrene polymer, with dynamic fragility $m \approx 100$. In Figure 4, we plot the measured and calculated $\alpha$- and SAP relaxation times for the relatively "strong" polyisobutylene (PIB) polymer, with dynamic fragility m ≈ 35. The $\alpha$-transition itself deviates from the Arrhenius behavior only very close to $T_g$, and the $\alpha$- and the SAP curves are nearly parallel. In fact, the difference between the SAP curves for PIB and PS is much less pronounced than for their $\alpha$-process curves.

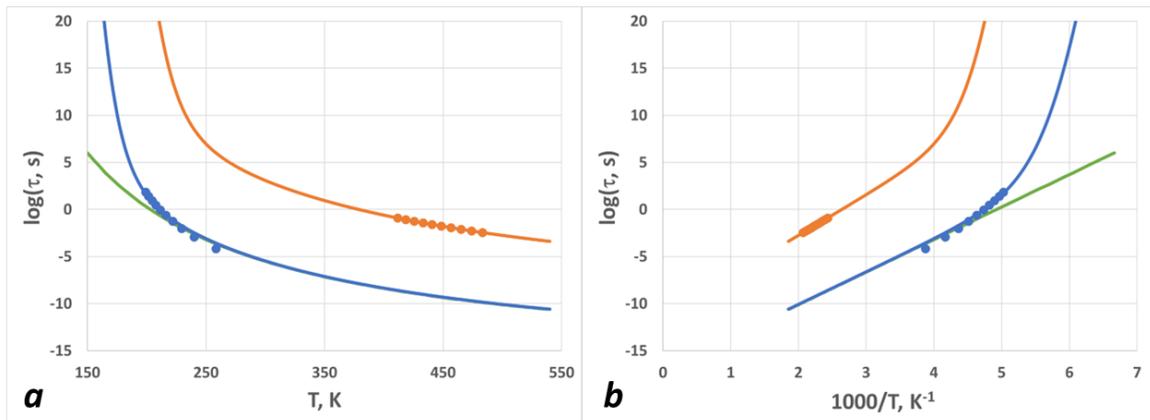

Figure 4. Same as Figure 3, but for PIB.

The data and model curves for the remaining eleven polymers are plotted in Supporting Information. We emphasize that the SAP model curves were generated with no adjustable parameters.

In Figure 5, we plot the Arrhenius time and Arrhenius temperatures for both $\alpha$- and SAP transitions of the thirteen polymers described in Table 1. The $\alpha$-transition parameters are depicted as orange circles and the SAP parameters are shown as blue circles. The straight line for the SAP corresponds to the Meyer-Neldel-type compensation. Obviously, no such correlation is seen for the $\alpha$-process.

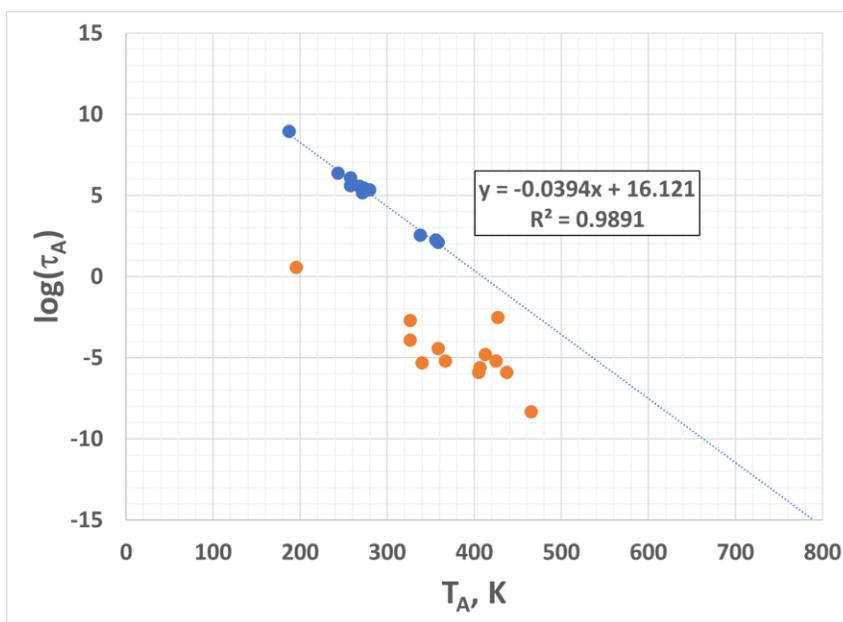

*Figure 5. Logarithm of the elementary time vs thermodynamic transition temperature for the $\alpha$-process (orange circles) and the SAP (blue circles). For the SAP, the Meyer-Nelder relationship holds, and there is a linear master curve shown as a trendline with a formula and the $R^2$-value.*

In Figure 6a-b, we depict the relationship between the two characteristic temperatures, $T_A$ (the Arrhenius temperature) and $T_g$ (the glass transition temperature, defined via $\log(\tau[T_g]) = 2$, where $\tau$ is either α-process or SAP characteristic time). The ratio $T_A/T_g$ is related to the dynamic fragility, *m*, monotonically increasing from ~0.86 to ~1.13 as the fragility is increased from ~20 (strong glass-formers) to ~150 (fragile glass-formers). With the exception of PNPMA and PIB, we find that $T_{A,\,SAP} \leq T_{A,\,\alpha}$, as seen in Figure 5a. Figure 5b shows that $T_{g,SAP}$ and $T_{g,\alpha}$ are generally scattered along the Y = X line, due to two opposite trends. On the one hand, the elementary time for the SAP is significantly higher than for the α-process, thus pushing $T_{g,SAP}$ upwards as well. On the other hand, the thermodynamic transition temperature ($T_x$ or $T_A$) is reduced, thus pushing $T_{g,SAP}$ downwards. The two effects partially cancel each other, resulting in $T_{g,SAP} \approx T_{g,\alpha}$.

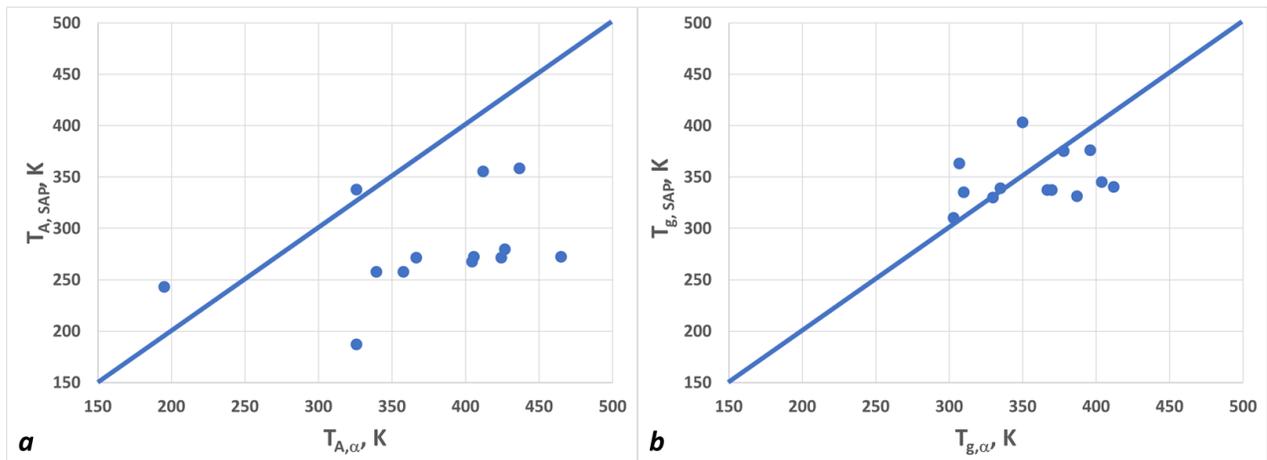

*Figure 6. The SAP vs. α-process transition temperatures: (a) Arrhenius temperature; (b) glass transition temperature. Blue lines correspond to Y = X equation.*

As already discussed by Napolitano and co-workers,[3–5] the elementary time ("the intercept") and the activation energy ("the slope") of the slow Arrhenius process are related via the linear Meyer-Neldel compensation law.[52,53] This is not the case for the α-process; if this were the case, all materials having the same $T_g$ would result in having the same fragility, which is obviously wrong. We discussed a simple model for the compensation law in the previous section – now let us apply the model to analyze the data.

In our earlier work,[34] it was found that the lower bound for the Arrhenius time in the limit of the high-fragility polymers is approximately given by, $\log(\tau_{A,\min}) \equiv \log(\tau^*) \approx -10$. In Figure 5, the blue line crosses the horizonal line Y = -10 (corresponding to the lower-bound elementary time) at $T_{A,\max} \equiv T^* \approx 660$ K. Let us suppose that the lower bound for the relaxation time corresponds to the upper bound for the coordination number where the system remains liquid. That would be given by $Z_{\max} = 2d = 6$ for the onset of jamming of spherical clusters.[57] This enables us to calculate the effective interaction energy $\varepsilon_{SAP} = \dfrac{2RT_{X,\max}}{aZ_{\max}} = \dfrac{2RT^*}{0.49 Z_{\max}} \simeq 3.7$ kJ/mol. This result is in excellent agreement with the calculations of Thoms and Napolitano,[3] where the same quantity was obtained via a direct analysis of the experimental data, namely from the slope of the correlation between the activation energy and the logarithm of the Arrhenius time.

The "dynamic slowing down" parameter $\lambda$ can be estimated as follows. In Figure 4, we see that the SAP line has the slope $A = 0.0394$ K$^{-1}$. This can be re-written as,

$$\log\left(\frac{\tau_{A,SAP}}{\tau^*}\right) = A(T^* - T_{A,SAP}) = (\log(\lambda))(N_{\max} - N)$$
$$= \frac{1}{2}(\log(\lambda)) N_0 (Z_{\max} - Z) = \frac{1}{2}(\log(\lambda)) N_0 \left(\frac{Z_{\max}}{T^*}\right)(T^* - T_{A,SAP})$$

(14)

leading to,

$$\log(\lambda) = \frac{2AT^*}{N_0 Z_{max}}$$

(15)

Thus, we obtain $\lambda \approx 1.52$. In other words, the slowdown per single contact point is fairly moderate (only 52%), but when amplified over the large number of contact points (~100 -- 300), it results in a very significant (apparent) elevation of the elementary time.

Next, we can calculate the apparent coordination number, Z, for each polymer for the α-process and the SAP. They are governed by the same equation,

$$Z = Z_{max} - \frac{2}{N_0 \log(\lambda)} \log\left(\frac{\tau_A}{\tau^*}\right) \quad (16)$$

Here, $\tau_A$ applies to both α-process and SAP.

The effective interaction energy, ε (in kJ/mol), is then given by,

$$\varepsilon = \frac{2RT_A}{0.49Z} \quad (17)$$

Based on the above, we can replot the α-process and SAP parameters in a different way, with the coordination number, Z, as the X-axis, and the interaction energy (in kJ/mol), ε, as the Y-axis, see Figure 7. Again, the blue dots correspond to SAP, and the orange dots to the α-process.

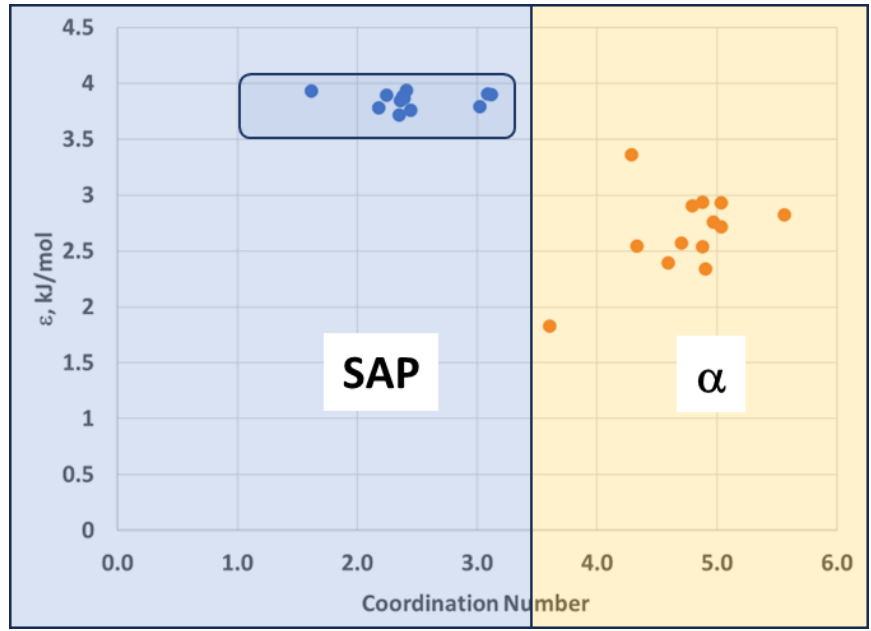

*Figure 7. Estimated bond interaction energy vs. coordination number. The vertical line at Z = 3.5 separates the SAP data (blue) from the α-process data (orange).*

Figure 7 thus interprets the experimental data based on the picture sketched in Figure 1. The α- and β-processes correspond to the relaxations of the segments or molecules. The elementary bond energy, $\varepsilon_\alpha$, is material-specific, ranging between 1.8 (for PIB) to 3.4 (for PTBS) kJ/mol. The coordination number $Z_\alpha$ ranges between 3.6 (for PIB) and 5.5 (for PC), depending on the material. The SAP process corresponds to the relaxation of the domains, and the "bond interaction energy" between domains, $\varepsilon_{SAP}$, is almost universal (at least for the polymer subset characterized here), $\varepsilon_{SAP} \approx 3.7$ kJ/mol. The SAP coordination number is relatively small, ranging from 1.8 to 3.2.

### *3.2. Discussion*

We propose a novel interpretation of the slow Arrhenius process (SAP), in which, through TS2 formalism, this mechanism is treated as the high-temperature $\alpha\beta$-process of the "clusters". The fluid of clusters, or the coarse-grained-particle (CGP) fluid, is expected to have a lower temperature of the thermodynamic transition, $T_x$, [28,34] as well as a lower fragility, as compared to the material itself. Hence, experimentally, SAP is seen as an Arrhenius process; we predict that is one were able to measure the same peak at lower temperatures and lower frequencies, it would be possible to see VFTH-like deviation from the Arrhenius behavior.

One interesting question is how thermodynamics of the CGP fluid differs from that of the material itself. The "general" SL-TS2 assumes that each material has two independent parameters, e.g., $T_A$ (Arrhenius temperature), and $\tau_A$ (Arrhenius time). In turn, $T_A$ depends on two factors – the number of neighbors, $Z$, and the van-der-Waals interaction parameter, $\varepsilon$. We propose that $Z$ is directly related to $\tau_A$ (eq 13). Thus, the general TS2 model can be reinterpreted in terms of two parameters, $\varepsilon$ and $Z$. In that case, we can immediately see that for the coarse-grained fluids (the slow Arrhenius process), the van-der-Waals parameter is fairly constant (it is also in agreement with the findings of Lipson et al.) We are still uncertain about the physical and molecular origins of this universality and whether it is specific to polymers and small organic molecules or can be observed in other (non-carbon-based) materials.

In our analysis, we hypothesized that the lower bound for the Arrhenius time is on the order of $10^{-10}$ s (corresponding to materials with high fragility $m \sim 150$). It has been known for a long time that the "absolute" lower bound on the material relaxation time is on the order of $10^{-13} - 10^{-14}$ s (see, e.g., Angell).[58] Thus, for the "fragile" materials, the Arrhenius region is relatively narrow (about 3-4 orders of magnitude), while for the "strong" materials, it can be quite wide (6-15 orders of magnitude). Obviously, the SAP materials fall into the second category. We stipulate that low fragility is associated with the low number of active contacts, somewhat consistent with the fact that low-coordination oxide networks are typically strong glass-formers, although this stipulation still needs to be tested further.

One other important point is about the potential limitations of our approach. The proposed theory is essentially a "mapping" from the Arrhenius equation to the "general TS2" equation. Both equations have two parameters, but the latter one describes a transition from the Arrhenius region to the WLF one (as well as the transition from WLF to the low-temperature Arrhenius that happens theoretically at times on the order of $10^{40} - 10^{60}$ s, well beyond our experimental capabilities to observe it). Thus, while we can always "generate" a TS2 extension from any Arrhenius equation, it may not always work. Specifically, if the TS2 model predicts the Arrhenius temperature to be too high, the predictions would fail. We found, for example, that among the polymers described by Thoms and Napolitano,[3] PBVCl is one example where general TS2 fails to capture the SAP dynamics; similarly, it fails to capture the $\alpha$-process for water. We also note that the potential super-Arrhenius behavior of the SAP relaxation time may or may not conform to the "general TS2" equation (or even VFTH-like behavior) if other interactions become important. Understanding the molecular mechanisms underpinning the TS2 universality and exceptions from it will be the topic of future research.

Overall, the proposed framework offers a new way of comprehensively modeling the slow Arrhenius process and therefore the phenomena determined by it (such as polymer adsorption), alongside the standard α relaxation. Future work should include further elucidating molecular mechanisms and origins of this "coarse graining by nature" and predicting the shapes and sizes of the "clusters" that give rise to SAP.

The above analysis dealt with the phenomenology of the SAP and the $\alpha$ transition; we now briefly discuss the physical origins of the cluster formation and the separate timescales.

### *3.3 Physical Origin of Cluster Formation and Timescale Separation*

The coarse-grained clusters invoked in the present framework should not be interpreted as permanent structural aggregates or thermodynamically distinct phases. Rather, they are dynamically emergent entities arising from the separation of relaxation timescales in glass-forming systems. As temperature approaches the glass transition, the dynamics become spatially heterogeneous and correlated over finite length scales. These correlated regions, often associated with cooperatively rearranging regions (CRRs),[20] represent domains within which molecular motions are strongly coupled.

Within the TS2 picture, the α-relaxation corresponds to the collective rearrangement of such domains. However, when the system is observed on timescales much longer than the local β-relaxation but still shorter than the slow Arrhenius process, a new effective dynamical hierarchy emerges. Specifically, for observation times satisfying

$$\tau_\beta \ll t \ll \tau_{SAP},$$

local motions and intra-domain relaxations have already equilibrated, while inter-domain rearrangements associated with the SAP have not yet occurred. In this intermediate time window, groups of correlated domains behave as effectively rigid units. The "clusters" introduced here are therefore dynamically defined coarse-grained particles: they are rigid not because their internal structure is frozen permanently, but because their internal relaxation is fast compared to the timescale of inter-cluster rearrangement.

This timescale separation naturally leads to hierarchical coarse-graining. At short times, the relevant degrees of freedom are molecular segments; at intermediate times, domains dominate; and at longer times, clusters of domains become the effective relaxing entities. The slow Arrhenius process can then be interpreted as the structural relaxation of this renormalized, coarse-grained fluid. In this sense, the SAP is not an independent relaxation mechanism, but the α-like relaxation of a system whose elementary units have been dynamically redefined by the underlying heterogeneity of the energy landscape.

The formation of dynamically correlated clusters may be understood as a consequence of collective constraints and connectivity. As temperature decreases, the number of effective strong contacts between domains increases, leading to transiently connected networks of domains whose collective rearrangement requires the cooperative displacement of many units. The activation barrier for such rearrangements scales with the number of effective contacts, which provides a natural explanation for the large Arrhenius activation energies associated with the SAP. Because the effective interaction energy at the coarse-grained level is approximately material-independent within the polymer systems examined, the resulting Arrhenius parameters exhibit Meyer–Neldel compensation behavior.

Importantly, this interpretation implies that the clusters are defined operationally by dynamical criteria rather than by static structural signatures. Their size and lifetime are determined by the competition between intra-domain relaxation and inter-domain rearrangement, and therefore depend on temperature and observation timescale. This perspective places the slow Arrhenius process within a broader framework of hierarchical relaxation in disordered materials and supports the view that glassy dynamics emerge from successive levels of dynamical coarse-graining.

## 4. Conclusions

We analyzed the slow Arrhenius process (SAP) using the framework of the "general" SL-TS2 theory. It is proposed that SAP represents the high-temperature region of the $\alpha$-process for the "coarse grained" fluid of clusters; the coarse-grained fluid has lower fragility and lower transition temperature relative to the material itself. Based on this analysis, we predict that the "slow Arrhenius process" would show an upward, VFTH-like deviation from the Arrhenius behavior at temperatures below the glass transition temperature. We also proposed the mechanism for the entropy-enthalpy compensation observed for the SAP – but not for the $\alpha$-process – the independence of the coarse-grained van-der-Waals energy on the material characteristics. Understanding molecular mechanisms of this independence and potential scenarios for its violation will be a subject of future studies.

Supporting Information for

# Modeling the Slow Arrhenius Process (SAP) in Polymers

Valeriy V. Ginzburg,[1,*] Oleg V. Gendelman,[2] Simone Napolitano,[3] Riccardo Casalini,[4] and Alessio Zaccone[5]

[1]Department of Chemical Engineering and Materials Science, Michigan State University, East Lansing, Michigan, USA 48824

[2]Faculty of Mechanical Engineering, Technion, Haifa 3200003, Israel

[3]Laboratory of Polymer and Soft Matter Dynamics, Experimental Soft Matter and Thermal Physics (EST), Universite libre de Bruxelles (ULB), Brussels 1050, Belgium

[4]Chemistry Division, Naval Research Laboratory, 4555 Overlook Avenue SW, Washington, D.C., USA 20375

[5]University of Milan, Department of Physics, via Celoria 16, 20133 Milano, Italy

*Corresponding author: ginzbur7@msu.edu

# Plots of the α- and SAP relaxation times for various polymers

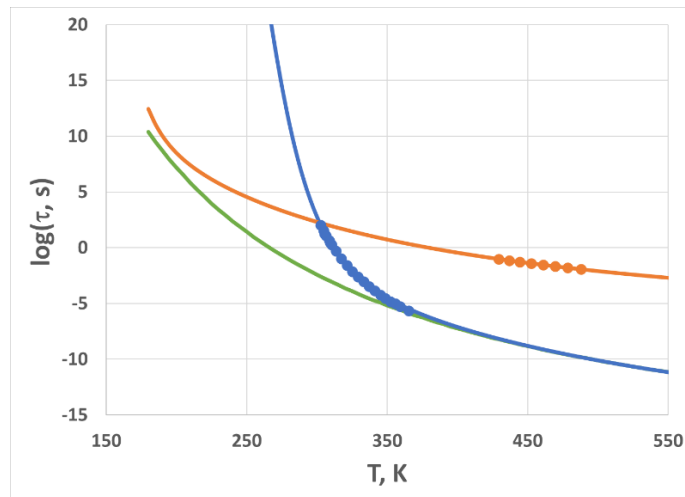

*Figure S 1. Experimental data (circles) and TS2 model fits for PVAc. Orange circles are the data for SAP relaxation times, and blue circles are the data for the α-relaxation times. Orange line is the TS2 model for SAP, blue line is the TS2 model for the α-process, and green line is the SAP model for the JG β-process.*

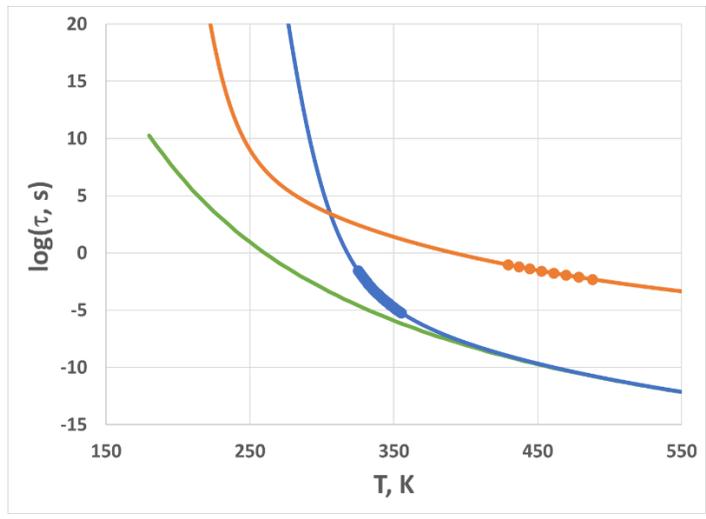

*Figure S 2. Same as Figure S1, but for PTBuA.*

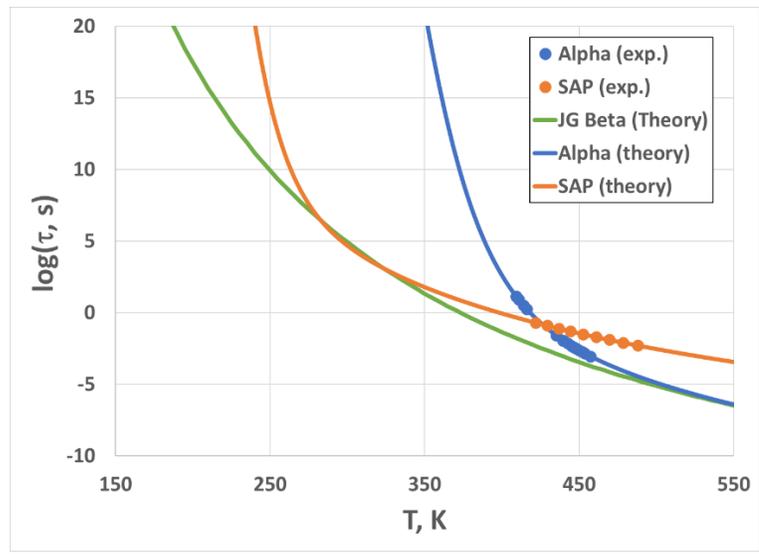

*Figure S 3. Same as Figure S1, but for PTBS.*

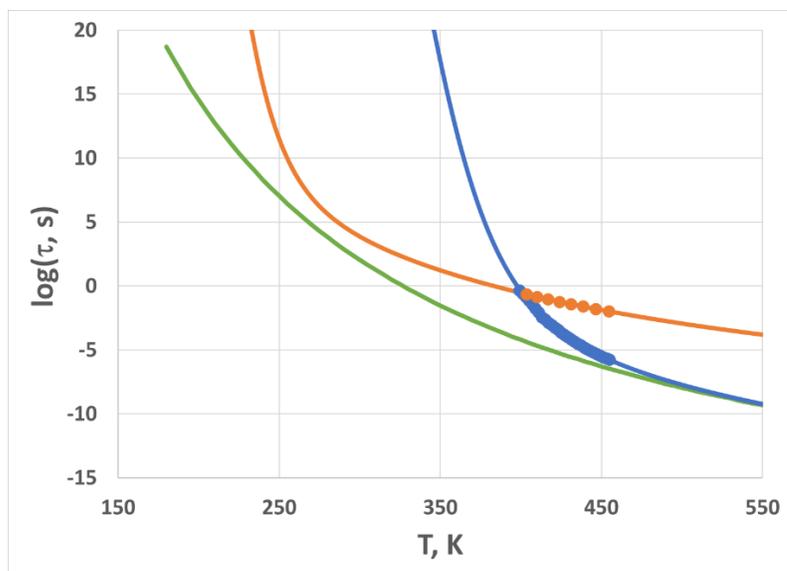

*Figure S 4. Same as Figure S1, but for PTBMA.*

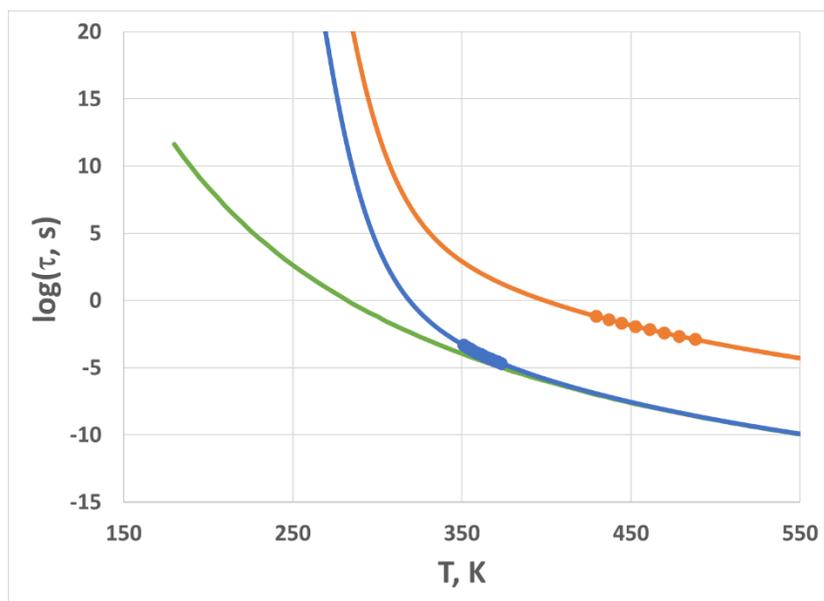

*Figure S 5. Same as Figure S1, but for PnPMA.*

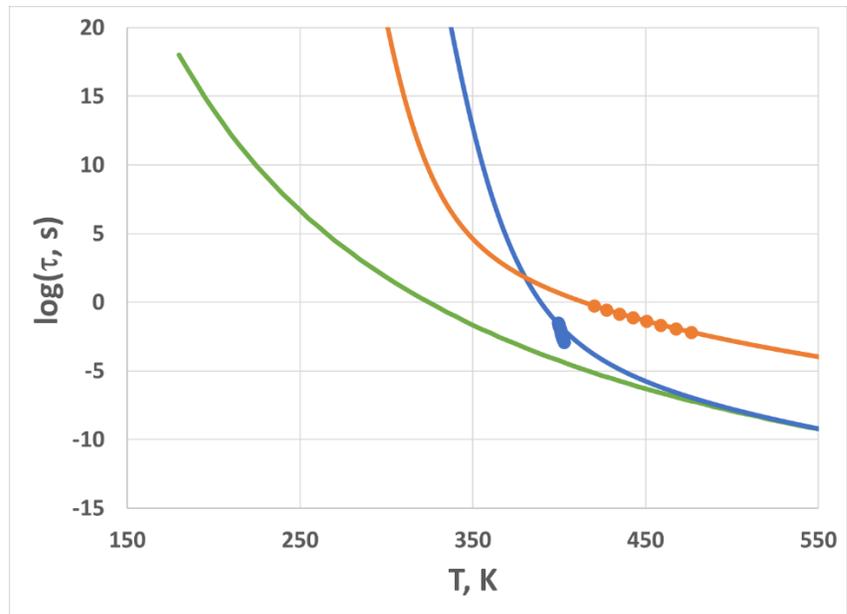

*Figure S 6. Same as Figure S1, but for PMMA.*

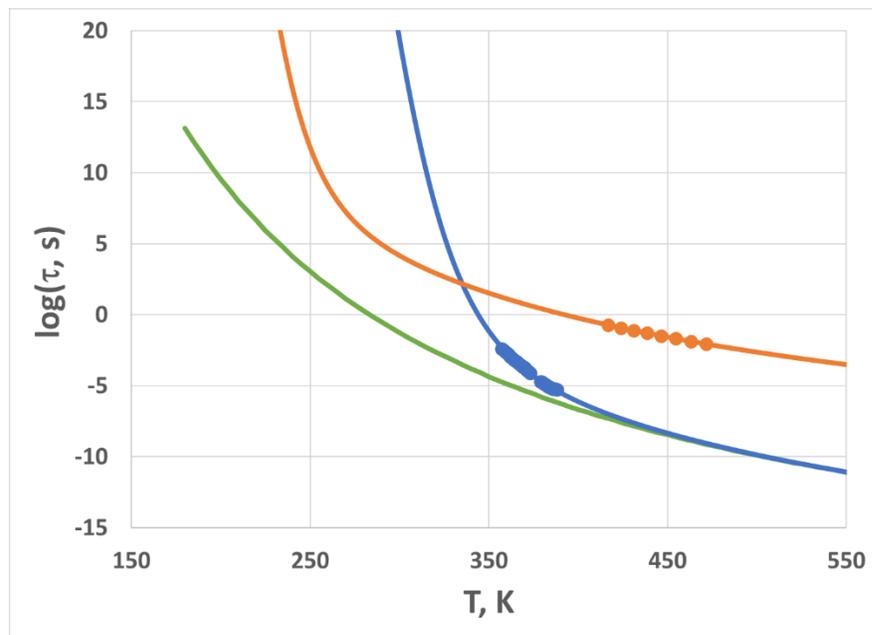

*Figure S 7. Same as Figure S1, but for PEMA.*

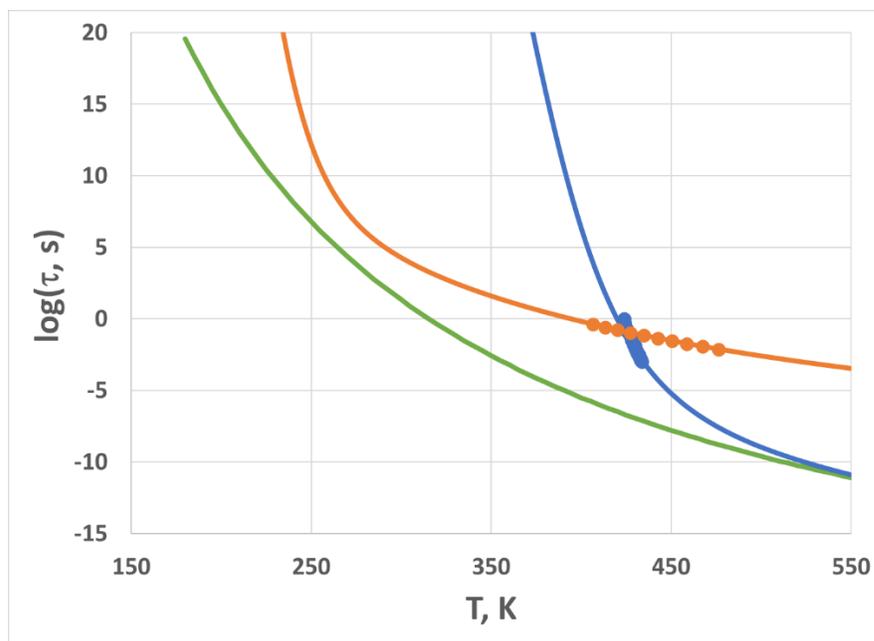

*Figure S 8. Same as Figure S1, but for PC.*

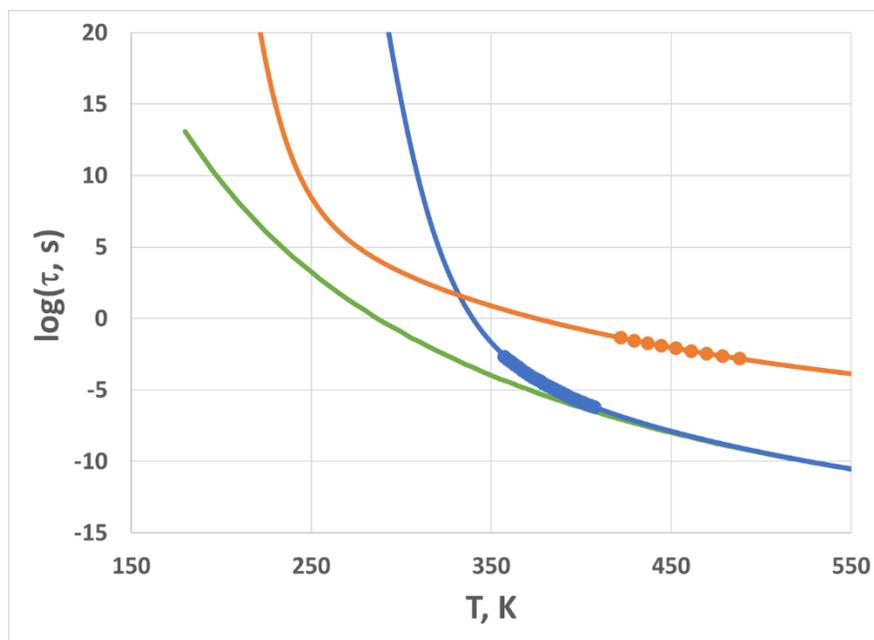

*Figure S 9. Same as Figure S1, but for PBzMA.*

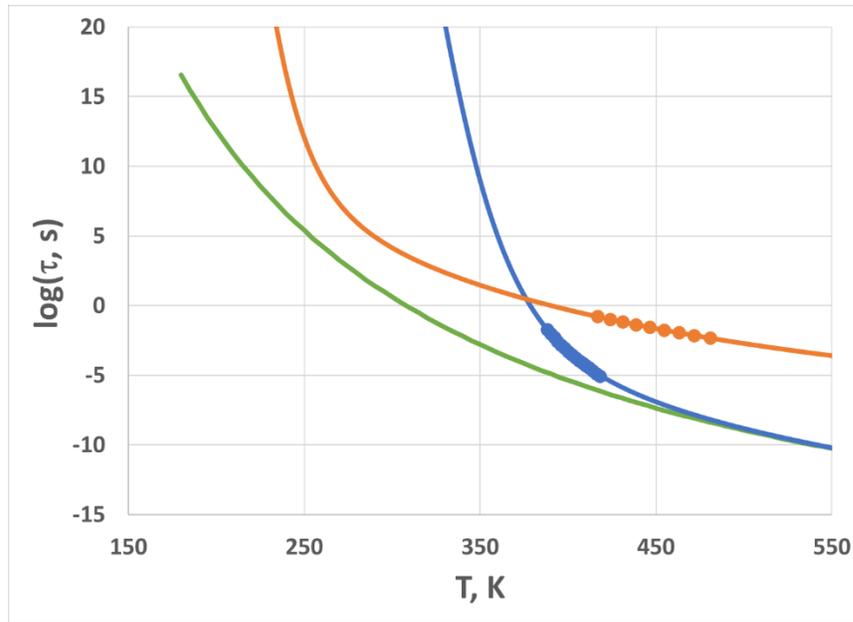

*Figure S 10. Same as Figure S1, but for P4MS.*

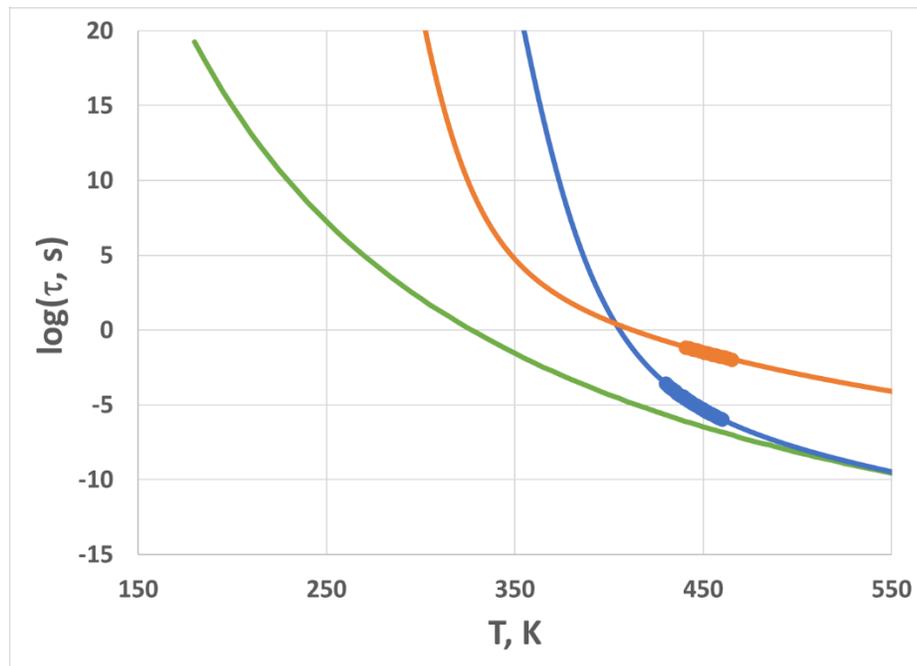

*Figure S 11. Same as Figure S1, but for P4ClS.*